\documentclass[twocolumn,showpacs,preprintnumbers,amsmath,nofootinbib]{revtex4}
\usepackage{graphics}

\begin{document}

\preprint{hep-th/0301129}
\preprint{FERMILAB-Pub-03/014-A}

\title{Spinors,   Inflation,  and  Non-Singular   Cyclic  Cosmologies}
\author{C.    Armend\'ariz-Pic\'on}  \email{armen@oddjob.uchicago.edu}
\affiliation{Enrico  Fermi  Institute,  Department  of  Astronomy  and
  Astrophysics, \\ University of Chicago.}  \author{Patrick B. Greene}
   \affiliation{ NASA/Fermilab Astrophysics Group, \\ Fermi National
        Accelerator Laboratory,\\ Batavia, IL 60510, USA}
\date{\today}

\begin{abstract}
  We consider toy cosmological models in which a {\em classical},
  homogeneous, spinor field provides a dominant or sub-dominant
  contribution to the energy-momentum tensor of a flat
  Friedmann-Robertson-Walker universe.  We find that, if such a
  field were to exist, appropriate  choices  of  the spinor
  self-interaction would  generate a  rich variety  of behaviors, quite
  different from their widely studied scalar  field counterparts. 
  We first discuss  solutions that incorporate a stage of cosmic inflation
  and  estimate  the primordial spectrum
  of density  perturbations seeded during  such a stage.  Inflation
  driven by  a spinor field turns out to be unappealing as it leads
  to a blue spectrum of perturbations and requires considerable
  fine-tuning of parameters.  We next find that, for simple, quartic spinor
  self-interactions, non-singular  cyclic cosmologies  exist with
  reasonable parameter choices.  These solutions might eventually be
  incorporated into a successful
  past- and future-eternal cosmological model  free  of  singularities.  
  In  an Appendix,  we  discuss  the classical treatment  of spinors and
  argue that certain quantum systems might be approximated in terms of
  such fields.
\end{abstract}


\maketitle

\section{Introduction}

At  least   since  the  advent   of  the  first   inflationary  models
\cite{Linde},  cosmologies  containing  classical scalar  fields  have
received  widespread  attention  in  the literature.   From  a  purely
phenomenological point of view,  such scalar fields are general enough
to accommodate a rich variety  of behaviors.  From a theoretical point
of  view, their invariable  appearance in  various theories  of nature
makes them natural candidates for cosmological applications.  In spite
of  these facts, one  might wonder  to what  extent scalar  fields are
singled  out  by the  former  considerations.   Could other  classical
homogeneous fields play a significant role in cosmology?

In  the present  paper, we  consider  the possibility
that {\em classical}, homogeneous,  spinor fields might play a role in
cosmology.   By a classical spinor
field, we simply mean a  set of four complex-valued spacetime functions
that transform  according to the spinor representation  of the Lorentz
group.   Although  the  existence   of  spin-$1/2$  fermions  is  both
theoretically  and experimentally undisputed,  these are  described by
{\em  quantum} spinor  fields. It  is unclear  when  fermionic quantum
fields  might be  consistently treated  as classical  spinors.   It is
generally held  that there exists  no classical limit  for fundamental
quantum Fermi  fields; however, one  can imagine classical  spinors as
arising  from  an effective  description  of  a  more complex  quantum
system.  We address possible justifications for the existence of classical
spinors in  an Appendix.  For the  bulk of this  paper, we  will simply
presuppose their existence.

We   find  that   classical  spinors   are   mathematically  perfectly
consistent.  Physically, one might object that spinors violate Lorentz
invariance  and isotropy.   Without being  explicit about  the precise
nature   of  Lorentz  transformations   in  a   general  gravitational
background, let us point out  that Lorentz invariance is broken in any
Friedmann-Robertson-Walker cosmology,  regardless of whether  a spinor
has a non-vanishing  expectation value or not.  On  the other hand, we
shall see that eventual violations of isotropy caused by the spinor do
not prevent  consistent solutions  of Einstein's equations,  and might
actually remain undetectable.

Compared  to  scalar  fields,  spinor  fields  have  attracted  little
attention in cosmology.  One of the first papers about the subject was
Taub's  study of  the Dirac  equation in  various  cosmological spaces
\cite{Taub}.  Brill and Wheeler  dealt with neutrinos in gravitational
fields \cite{BrillWheeler}.  The quantization  of a spinor field in an
expanding  universe was  considered by  Parker \cite{Parker},  and the
quantization  of  gravity  coupled   to  a  spinor  was  addressed  in
\cite{IshamNelson,ChristodoulakisZanelli}.  Explicit  solutions of the
Dirac  equation in an  open Friedmann-Robertson-Walker  spacetime have
been considered  in \cite{KovalyovLegare}.  Solutions  of the Einstein
equations  coupled to  a spinor  in Bianchi  Type I  spaces  have been
extensively studied  by Saha  and Shikin \cite{SahaShikin}.

The   structure  of  this   paper  is   the  following.    In  Section
\ref{sec:formalism} we discuss how to couple a spinor to gravity.  The
reader  familiar with the  formalism might  want to  skip to  the next
section  and eventually  refer back  for notational  details.  Section
\ref{sec:solutions}  deals with the  basic cosmological  equations and
general  solutions in  terms of  an arbitrary  spinor self-interaction
term.  In  Section \ref{sec:inflation} we study inflation  driven by a
spinor  field   and  compute   the  spectrum  of   primordial  density
perturbations.   Section Section  \ref{sec:cyclic} presents  a cyclic,
non-singular  model of  the  universe that  critically  relies on  the
properties    of    a    spinor    field.    Finally,    in    Section
\ref{sec:conclusions}   we  summarize   our  results   and   draw  our
conclusions. The  Appendix comments on  the meaning and  properties of
classical spinors.

\section{Formalism}\label{sec:formalism}
In this  section we briefly  review how a  spinor field is  coupled to
gravity.  For complete discussions about spinors in curved spacetimes,
see \cite{Weinberg,BirrellDavies,GSW}.

Because   the  group   of  diffeomorphisms   does  not   admit  spinor
representations,  in  order to  couple  a  spinor  to gravitation  one
introduces  the  Lorentz  group   (which  does  actually  have  spinor
representations)  as a  local  symmetry group  of  the theory.   Under
diffeomorphisms $x^\mu\to \tilde{x}^\mu(x^\nu)$,  a spinor $\psi$ is a
scalar,  $\psi\to  \tilde{\psi}=\psi$,   but  under  a  local  Lorentz
transformation with  parameters $\lambda_{ab}(x)$ a  spinor transforms
according to
\begin{equation}\label{eq:lltspinor}
  \psi\to \tilde{\psi}=\exp\left[\frac{1}{2}\lambda_{ab}(x)
\Sigma^{ab}\right]\psi,
\end{equation}
where    $\Sigma^{ab}\equiv\frac{1}{4}[\gamma^a,\gamma^b]$   are   the
generators of the spinor representation  of the Lorentz group, and the
$4\times4$   matrices   $\gamma^a$   satisfy  the   Clifford   algebra
$\{\gamma^a,\gamma^b\}=2\eta^{ab}$.   We shall choose  the Dirac-Pauli
representation
\begin{equation}
  \gamma^0=\left(\begin{array}{cc} 
      1 & 0 \\ 0 & -1 \end{array}\right)\, \quad
  \gamma^i=\left(\begin{array}{cc} 
0 & \sigma^i \\ -\sigma^i & 0 \end{array}\right),
\end{equation}
where the  $\sigma^i$ are the conventional $2\times2$  Pauli matrices. 
Then,  $\gamma^0=(\gamma^0)^\dag$ is  Hermitean, and  the  $\gamma^i =
-(\gamma^i)^\dag$ are anti-Hermitean.  For later convenience, we shall
define  the  additional  (Hermitean)  gamma matrix  $\gamma_5\equiv  i
\gamma_0\gamma_1\gamma_2\gamma_3$.

A fermion  is coupled  to gravitation with  the aid of  a ``vierbein''
$e^\mu{}_a$, a  set of four  contravariant vector fields  that satisfy
the orthonormality condition
\begin{equation}\label{eq:vierbein}
  g_{\mu\nu}   e^\mu{}_a  e^\nu{}_b=\eta_{ab} \, ,
\end{equation}
where  $g_{\mu\nu}$ is  the spacetime  metric and  $\eta_{ab}$  is the
Minkowski  metric $\eta_{ab}=\text{diag}(1,-1,-1,-1)$.   Latin indices
enumerate  each of  the vectors  in the  vierbein while  Greek indices
enumerate  the  spacetime  components   of  each  of  these  vectors.  
Spacetime  and  Lorentz  indices  are  raised  and  lowered  with  the
spacetime  and Minkowski metrics  respectively, leading  to associated
sets of vectors such as $e_{a \mu}$ and $e_\mu{}^a$.

Local   Lorentz   transformations    $\Lambda(x)$   are   just   local
``reshufflings'' of the vierbein vectors
\begin{equation}\label{eq:lltvierbein}
  e^\mu{}_a\to \tilde{e}^\mu{}_a=\Lambda_a{}^b e^\mu{}_b,
\end{equation}
that preserve the orthonormality relation~(\ref{eq:vierbein}) at each
point.  Thus, the spacetime metric  only determines the vierbein up to
such local Lorentz transformations.   For this reason, one must ensure
that any Lagrangian  formed with the aid of  the vierbein is invariant
under the Lorentz  group acting as a local  symmetry.  Invariant terms
containing  derivatives of  a spinor  can be  constructed  through the
covariant derivative
\begin{equation}
D_\mu \psi=(\partial_\mu + \Omega_\mu)\psi,
\end{equation}
which transforms as a  (covariant) vector under diffeomorphisms and as
a spinor under local  Lorentz transformations.  The $4\times 4$ matrix
$\Omega_{\mu}$ is the spin connection
\begin{equation}\label{eq:spinconnection}
  \Omega_\mu=\frac{1}{2} \omega_{\mu ab}\Sigma^{ab},\quad
\omega_{\mu ab}=e^\nu{}_a \nabla_\mu e_\nu{}_b  \, .
\end{equation}
The coefficients  $\omega_{\mu ab}$ are  the Ricci rotation  (or spin)
coefficients.

The vierbein and  the flat-space gamma matrices allow  one to define a
new set of gamma matrices
\begin{equation}
\Gamma^\mu\equiv e^\mu{}_a \gamma^a
\end{equation} 
that  satisfy\footnote{  Note  that,  while the  $\gamma^a$'s  do  not
  transform  under local  Lorentz transformations  or diffeomorphisms,
  the $\Gamma^\mu$'s do.  They inherit their transformation properties
  from          the          vierbein.}           the          algebra
$\{\Gamma^\mu,\Gamma^\nu\}=2g^{\mu\nu}$.  These  can be used  to write
down  a generalization  of  the  Dirac action  in  a curved  spacetime
background,
\begin{equation}\label{eq:Lagrangian}
  S_\psi=\int d^4x \,e \,
  \left[\frac{i}{2}\left(\bar{\psi}\Gamma^\mu D_\mu\psi
    -D_\mu \bar{\psi}\Gamma^\mu \psi\right)-V\right],
\end{equation}
which  we  have  written in  a  symmetrized  form.  Here, $e$  is  the
determinant  of  the  vierbein  $e_\mu{}^a$,  and  the  Dirac  adjoint
$\bar{\psi}$  is   given  by  $\psi^\dag   \gamma^0$.   The  covariant
derivative      acting     on      the      adjoint     is      $D_\mu
\bar{\psi}=\partial_\mu\bar{\psi}-\bar{\psi}\,\Omega_\mu$.     By   an
integration by  parts, the kinetic term  of the spinor can  be cast in
the  ``conventional'' form  $i\bar{\psi}\Gamma^\mu  D_\mu \psi$.   The
term $V$  stands for any  scalar function of $\psi$,  $\bar{\psi}$ and
possibly additional matter  fields.  When a particular form  of $V$ is
later  needed, we  will assume  that $V$  only depends  on  the scalar
bilinear $\bar{\psi}\psi$.   It turns out that this  choice is general
enough for our purposes.  More  general interactions in Bianchi Type I
spacetimes   have   been   considered   in  the   series   of   papers
\cite{SahaShikin}.

The  Lagrangian  (\ref{eq:Lagrangian})  describes  how the  spinor  is
coupled to gravity,  but it does not specify the  dynamics of gravity. 
We shall  assume that the  latter is governed by  the Einstein-Hilbert
action.   Hence, we  consider a  spinor minimally  coupled  to general
relativity,
\begin{equation}\label{eq:action}
  S=S_\psi+S_m-\frac{1}{6}\int d^4x \sqrt{-g}R,
\end{equation}
where  $R$  is  the  scalar  curvature,  $S_\psi$  is  given  by  Eq.  
(\ref{eq:Lagrangian})  and $S_m$  describes additional  matter fields,
such  as  scalar fields  or  gauge  fields.   The symmetries  we  have
postulated  up to  now, diffeomorphism  and local  Lorentz invariance,
certainly allow  for the presence of  additional terms in  the action. 
For  example, we  could  have  written down  a  non-minimal term  like
$\bar{\psi}\psi R$. However,  as we are going to  see, in an expanding
universe  $\bar{\psi}\psi$  decays  at  least  as  fast  as  $1/a^3$.  
Therefore, during  cosmic expansion such  a term would  quickly become
negligible. This  situation is  in sharp contrast  with the case  of a
scalar field $\phi$, where {\it a priori} there is no reason to expect
a  term  proportional  to  $\phi  R$ to  be  negligible  (see  however
\cite{Carroll,DamourPolyakov}).    Eventually  this   fact   could  be
relevant  in models  where  a  spinor field  drives  late time  cosmic
acceleration.

Varying  the action  (\ref{eq:action})  with respect  to the  vierbein
$e^\mu{}_a$ leads to the Einstein equations
\begin{equation}\label{eq:Einstein}
G_{\mu\nu}=3 T_{\mu\nu},
\end{equation}
where  the  energy  momentum  tensor  $T_{\mu\nu}$  is  given  by  the
variation of the matter action,
\begin{equation}\label{eq:defemt}
  T_{\mu\nu}=\frac{e_{\mu a}}{e}\frac{\delta S}
  {\delta e^\nu{}_a}.
\end{equation}
Note that  we work in units where  $8\pi G/3=\hbar=c=1$.  Substituting
the  action  (\ref{eq:Lagrangian})   into  Eq.   (\ref{eq:defemt})  we
obtain, after an  integration by parts, the energy  momentum tensor of
the spinor (on-shell),
\begin{equation}\label{eq:emt}
  {}^{(\psi)} T_{\mu\nu}=\frac{i}{2}\left[\bar{\psi}\Gamma_{(\mu}
    D_{\nu)} \psi-D_{(\nu} \bar{\psi} \Gamma_{\mu)} \psi\right]
  - g_{\mu\nu}L_\psi \, .
\end{equation}
We have used a relation that follows from the Lorentz invariance
of the spinor Lagrangian,
\begin{eqnarray}
  \lefteqn{D_\mu(\bar{\psi}\{\Gamma^\mu,\Sigma^{\rho\sigma}\}\psi)=} 
  \nonumber \\
&=\bar{\psi}\Gamma^\rho D^\sigma\psi-\bar{\psi}\Gamma^\sigma D^\rho\psi
-(D^\sigma \bar{\psi})\Gamma^\rho\psi+(D^\rho\bar{\psi})\Gamma^\sigma \psi,
\nonumber \\
\end{eqnarray}
to rewrite  the result  of the spinor  variation.  On the  other hand,
varying the action  with respect to the field  $\bar{\psi}$ yields the
equation  of motion  of  the  spinor, a  generalization  of the  Dirac
equation to a curved spacetime,
\begin{equation}\label{eq:Dirac}
  i \Gamma^\mu D_\mu \psi-\frac{\partial V}{\partial \bar{\psi}}=0 \, .
\end{equation}
If the action is real, the variation of
the action with respect to $\psi$ yields the adjoint of the previous
equation.

\section{Cosmological solutions}\label{sec:solutions}
Because we  are interested  in cosmology, in  this paper we  deal with
homogeneous and isotropic  FRW spacetimes.  Current observations favor
a flat universe  \cite{flat}, so we assume the  spacetime metric to be
spatially flat,
\begin{equation}\label{eq:metric}
  ds^2=dt^2-a^2(t)\, d\vec{x}^2.
\end{equation}

For these isotropic solutions of  the Einstein equations to exist, the
energy-momentum  tensor of  the  spinor must  be  compatible with  the
symmetries of the metric  (\ref{eq:metric}), homogeneity and isotropy. 
At  this  point,   note  that  homogeneity  of  a   spinor  is  not  a
gauge-invariant   concept;   by   a   local   Lorentz   transformation
(\ref{eq:lltspinor}), it is always possible to transform a homogeneous
(space-independent)    spinor   $\psi(t)$   into    an   inhomogeneous
(space-dependent) one $\tilde{\psi}(t,\vec{x})$.   We are going to look
for spinor  solutions of the Dirac  equation that can be  written as a
gauge-transformed  homogeneous spinor.   If  that is  the case,  there
exists a  vierbein where  the Dirac equation  allows space-independent
solutions.  In our  case such a vierbein is  given by\footnote{In open
  or closed  FRW universes, one  can construct a vierbein  that allows
  homogeneous  spinor solutions.   These are  formed from  the Killing
  vectors of  the homogeneous 3-dimensional  spaces of Bianchi  Type V
  and Type IX, respectively.}
\begin{equation}\label{eq:killinggauge}
  e^\mu{}_0 =\delta^\mu{}_0 \quad ,
  e^\mu{}_i=\frac{1}{a}\delta^\mu{}_i.
\end{equation}
In  the gauge  (\ref{eq:killinggauge})  the equation  of  motion of  a
space-independent spinor (\ref{eq:Dirac}) reads
\begin{equation}\label{eq:FRWDirac}
  \dot{\psi}+\frac{3}{2}H \psi+i \gamma^0 V' \psi=0,
\end{equation}
where  a  dot  (\,$\dot{}$\,)  denotes  a  time  derivative,  a  prime
(\,$'$\,) denotes  a derivative with respect  to $\bar{\psi}\psi$, and
$H=d(\log  a)/dt$ is  the Hubble  parameter.  The  equation manifestly
admits space-independent solutions,  and hence, spinor observables like
the energy momentum tensor are also homogeneous.

One  should  also  verify  whether  spinors are  compatible  with  the
isotropy of the FRW  metric.  The ${}_0{}^i$ Einstein  Eqs.  $0\equiv
G_0{}^i=T_0{}^i$ are  satisfied only  if $T_0{}^i$ vanishes.   This is
possible for conventional matter forms (perfect fluids and homogeneous
scalars),  but it  is not  generally true  for a  spinor. In  fact, in
spatially open or closed universes,  it is not possible to satisfy the
constraint    $T_0{}^i=0$     unless    $\bar{\psi}\psi$    is    zero
\cite{IshamNelson,  ChristodoulakisZanelli}.    In  a  spatially  flat
universe   however,  the   equation   of  motion   (\ref{eq:FRWDirac})
automatically implies the vanishing of $T_0{}^i$, so that the presence
of the spinor is consistent with the isotropy of the metric.

A   convenient   combination   of   the  remaining   Einstein   Eqs.   
(\ref{eq:Einstein}), the  ${}_0{}^0$ and the  ${}_i{}^j$, involves the
energy  density  $\rho_k$ and  the  pressure  $p_k$  of the  different
constituents of the universe,
\begin{eqnarray}
  H^2&=&\rho_{\psi}+\rho_{m}\label{eq:friedmann} \\
  \ddot{a}&=&
  -\frac{1}{2}[\rho_{\psi}+\rho_m+3(p_{\psi}+p_m)]a \label{eq:dotdota}.
\end{eqnarray}
Here,  the sub-index  $\psi$ stands  for the  spinor and  $m$  for any
additional matter, such as dust,  radiation, or even dark energy.  The
spinor's energy  density and pressure  are given by  the corresponding
components   of    the   energy   momentum    tensor   (\ref{eq:emt}),
\begin{eqnarray}
  \rho_\psi&\equiv&{}^{(\psi)}T_0{}^0=V \label{eq:rho}\\
  p_\psi&\equiv&{}-{}^{(\psi)}T_i{}^i=V'\bar{\psi}\psi-V. \label{eq:p}
\end{eqnarray}
The  equation of  state of  the spinor  $w_\psi$ is  the ratio  of its
pressure to energy density, and hence, it is given by
\begin{equation}\label{eq:w}
  w_\psi\equiv\frac{p_\psi}{\rho_\psi}=\frac{V' \bar{\psi}{\psi}-V}{V}.
\end{equation}
The  equation of state  is not  restricted to  be within  the interval
$-1\leq   w   \leq   1$.    For  a   conventional   massive   fermion,
$V=m\bar{\psi}\psi$, the equation of state agrees with that of a fluid
of dust, $w_\psi=0$.  For more  general choices of $V$, $w_{\psi}$ may
acquire any real value.

It is possible to directly  integrate the spinor equation of motion if
$V$ only depends on  $\bar{\psi}{\psi}$.  It follows directly from the
Dirac equation (\ref{eq:FRWDirac}) that
\begin{equation}\label{eq:dilution}
  \bar{\psi}{\psi}=\frac{A}{a^3},
\end{equation}
where $A$  is a time-independent  constant.  Note that this  result is
valid for any time dependence  of the background geometry, $a(t)$, and
thus,  is valid  regardless of  the dominant  energy component  of the
universe.  In  an expanding universe, the  value of $\bar{\psi}{\psi}$
monotonically decreases,  whereas in a contracting  universe the value
of  $\bar{\psi}{\psi}$ monotonically  increases.  These  facts  do not
imply however that  the energy density of the  spinor follows the same
behavior.  The Dirac equation can be cast as a continuity equation
\begin{equation}\label{eq:continuity}
  \dot{\rho}+3H\rho (1+w)=0.
\end{equation}
Integrating   Eq.   (\ref{eq:continuity})   or  directly   from   Eq.  
(\ref{eq:dilution}) it  is possible to find $\rho_\psi$  as a function
of the scale factor $a$ for arbitrarily given $V(\bar{\psi}\psi)$,
\begin{equation}
  \rho_\psi=V\big|_{\bar{\psi}\psi=A/a^3}.
\end{equation}
Conversely, given an arbitrary  function $\rho_\psi(a)$ one can always
find a  $V(\bar{\psi}\psi)$ such that  $\rho(a)$ is a solution  of the
equation of motion (\ref{eq:continuity}),
\begin{equation}
  V(\bar{\psi}\psi)
  =\rho_\psi\big|_{a=(A/ \bar{\psi}\psi)^{1/3}}.
\end{equation} 
In conclusion, a spinor field  can accommodate any desired behavior of
its energy density by an appropriate choice of $V$. In that respect, a
spinor  field  is  completely   different  from  a  scalar  field.   A
(canonical) scalar field  cannot violate the null\footnote{$\rho+p\geq
  0$} energy condition \cite{Wald}, whereas a spinor field can violate
any  desired---weak\footnote{$\rho+p\geq 0$  and $\rho\geq  0$}, null,
strong\footnote{$\rho+p\geq    0$     and    $\rho+3p\geq    0$}    or
dominant\footnote{$\rho\geq  |p|$}---one.   Even non-canonical  scalar
fields---``k-fields''  \cite{k-field}---cannot reproduce  the behavior
of  a  spinor.   In the  former  there  are  barriers that  prevent  a
transition from  $\rho+p>0$ to  $\rho+p<0$, whereas such  barriers are
nonexistent for spinors.  The converse  is however true.  A spinor can
reproduce the behavior of a scalar field.  In particular, it can drive
inflation and late time cosmic acceleration.

Although these solutions of Einstein equations sourced by a spinor are
perfectly valid and consistent, they might break isotropy.  By that we
mean  that the spatial  components of  certain vector  quantities that
involve  the spinor  do not  necessarily  vanish, and  hence, are  not
invariant under  spatial rotations.  For  instance, it turns  out that
for   non-trivial   homogeneous   solutions   of   the   Dirac   Eq.   
(\ref{eq:FRWDirac}) the spatial  components of the vector $j^\mu\equiv
\bar{\psi}\Gamma^\mu\psi$  are generically  non-zero.   If the  action
(\ref{eq:action}) does not include  a coupling between $j^\mu$ and any
other  observable   vector  quantity,   such  a  violation   would  be
undetectable.   On the  other hand,  if  the action  contained such  a
coupling, there still  exist some spinors for which  $j^i=0$, such as,
for instance,
\begin{equation}\label{eq:isotropicspinor}
  \psi=(\psi_1,0,0,0),
\end{equation}
Note that this  form of the spinor is compatible  with the equation of
motion  (\ref{eq:FRWDirac}).  In other  cases, no  choice of  a spinor
prevents  isotropy violations.   There is  no non-trivial  spinor such
that  the   ``pseudo-vector''  $\bar{\psi}\gamma_5\gamma^\mu\psi$  has
non-vanishing        spatial       components       \cite{IshamNelson,
  ChristodoulakisZanelli}.  But again, if there is no coupling between
the latter vector and any  other observable component (say, because of
parity conservation), such a violation would remain undetectable.

Although in this section we  have mainly assumed that $V$ only depends
on the scalar bilinear $\bar{\psi}{\psi}$,  some of the results can be
easily generalized for rather arbitrary choices of $V$.  Consider, for
example,  any $V$ that  is invariant  under the  global transformation
$\psi\to e^{i\alpha}  \psi$, for arbitrary constant  $\alpha$.  Such a
symmetry means that  the $\psi$ flavor is conserved,  and hence, there
is  a  conserved  current $\nabla_\mu(\bar{\psi}\Gamma^\mu  \psi)=0$.  
Then, for a homogeneous spinor
\begin{equation}
  \bar{\psi}\gamma^0 \psi=\frac{\tilde{A}}{a^3},
\end{equation}
which  already suggests that  Eq.  (\ref{eq:dilution})  is not  just a
consequence of our  choice of $V$. In fact,  writing down the 4-spinor
in terms of two 2-spinors, $\psi=(u, v)$, it follows from the identity
\begin{eqnarray}
  (\bar{\psi}\psi)^2+(i\bar{\psi}\gamma_5\psi)^2
  +(\bar{\psi}\gamma^0\gamma_5\psi)^2 = \nonumber \\ 
  (\bar{\psi}\gamma^0 \psi)^2-4\left[(u^\dag u)(v^\dag v)
  -(u^\dag v)( v^\dag u)\right] 
\end{eqnarray}
and the Cauchy-Schwarz inequality that
\begin{equation}
  (\bar{\psi}\psi)^2+(i\bar{\psi}\gamma_5\psi)^2+
  (\bar{\psi}\gamma^0\gamma_5\psi)^2\leq
  \frac{\tilde{A}^2}{a^6}.
\end{equation}
Therefore, spinor bilinears generically  decay during the expansion of
the  universe,  without regard  to  the  precise  form of  the  spinor
interaction.  In  particular, in an  expanding universe, there  are no
non-trivial solutions of the  spinor equations of motion with constant
$\psi$.

\section{Inflation}\label{sec:inflation}
In this section, we investigate the possibility that a classical
spinor field could drive inflation.
A sufficiently  long stage of inflation \cite{Linde}  explains many of
the features  of our universe  that remain unexplained  otherwise (see
\cite{WaldHollands} for  diverging claims).  Nevertheless,  the nature
of the component  that was responsible for inflation  remains unknown. 
Most inflationary scenarios rely on a homogeneous scalar field rolling
down an appropriate potential; however,  at present there is no direct
experimental evidence  for the existence of  fundamental scalar fields
in nature. Hence,  a  natural
question  is whether  a  another type of field  could  have driven  a stage  of
inflation in the  early universe.  Inflation driven by  a vector field
has been  considered by  Ford \cite{Ford}, and  inflation driven  by a
spinning fluid has been discussed by Obukhov \cite{Obukhov}.

\subsection{Background}

By definition,  inflation is a  stage of accelerated expansion  of the
universe, $\ddot{a}>0$.   It follows from  Eq. (\ref{eq:dotdota}) that
any component  driving inflation has  an equation of state  that obeys
$w<  -1/3$.  (We  assume  $\rho$  to be  positive.)   Three  types  of
inflation are mainly considered in the literature: pole-like inflation
with $w<-1$,  de Sitter inflation with $w=-1$  and power-law inflation
with $-1<w<-1/3$.   It is easy  to verify from Eq.   (\ref{eq:w}) that
inflation (or  simply expansion) with  constant equation of  state $w$
results from a ``potential''
\begin{equation}\label{eq:power-potential}
  V=(\bar{\psi}\psi)^{1+w}.
\end{equation}
If  $w<-1$ the  expansion runs  into  a future  singularity, while  if
$w>-1$ the expansion runs into  a past singularity.
However,
by an appropriate choice of $V$, the universe
could  pole-like inflate  in the  past and  power-like inflate  in the
future.

If $w=-1$, the formula  (\ref{eq:power-potential}) implies that $V$ is
constant, as  for a  cosmological term.  It  is possible to  relax the
condition on  the function  $V$ by  looking for a  stage of  nearly de
Sitter  inflation, $w\approx-1$.  In  terms of  the function  $V$, the
condition for nearly de Sitter inflation is
\begin{equation}\label{eq:slow-roll}
  \left|\frac{d\log V}{d\log \bar{\psi}\psi}\right|\ll 1.
\end{equation}
Note that  the latter condition  alone suffices to guarantee  quasi de
Sitter  inflation.    This  is   to  be  compared   with  conventional
scalar-field  driven  inflation, where  two  slow-roll conditions  are
needed.  In general, any $V$ that asymptotes to a positive constant at
large   $\bar{\psi}\psi$  satisfies   Eq.    (\ref{eq:slow-roll}).  
Examples  of  such functions  $V$  are $\log  [1+(\bar{\psi}\psi)^n]$,
$\tanh                    n\cdot\bar{\psi}\psi$                    and
$(\bar{\psi}\psi)^n/(1+\bar{\psi}\psi)^n$  for arbitrary  positive $n$
and sufficiently large $\bar{\psi}\psi$.

\begin{figure}
    \includegraphics{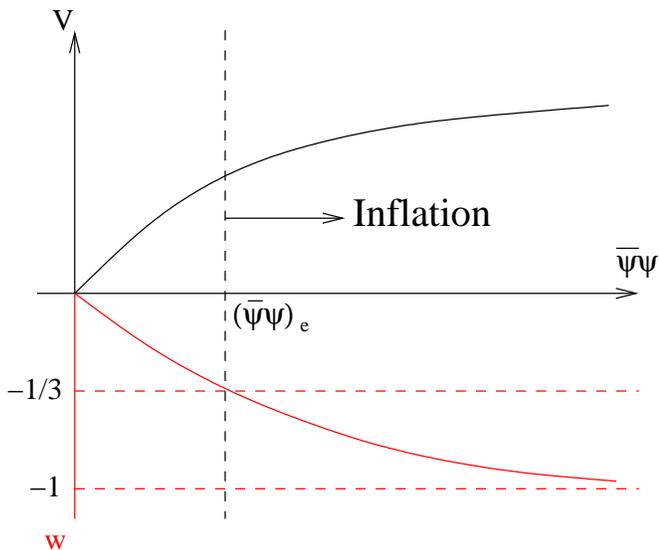}
  \caption
  {A  plot  of  a  generic  interaction that  yields  inflation.   The
    corresponding equation of state is  also shown in the diagram. For
    large  values of $\bar{\psi}\psi$,  the interaction  is flat,  Eq. 
    (\ref{eq:slow-roll}), allowing nearly de Sitter inflation . At the
    critical value  $(\bar{\psi}\psi)$ the equation  of states reaches
    $-1/3$ and inflation ceases to be possible.\label{fig:inflation}}
\end{figure}

Although  nothing  prevents a  spinor  field  from driving  inflation,
certain facts make this possibility unappealing.  Inflation solves the
homogeneity problem  if it  lasts for about  60 $e$-foldings.   Let us
assume     that     $V$      is     such     that     $w<-1/3$     for
$\bar{\psi}\psi>(\bar{\psi}\psi)_e$       and       $w=-1/3$       for
$\bar{\psi}\psi=(\bar{\psi}\psi)_e$  (Fig.  \ref{fig:inflation}).  The
end  of inflation  is  determined by  $(\bar{\psi}\psi)_e$, the  point
where the equation of state $w$ crosses the ``critical'' value $-1/3$.
For    instance,    for    $V=(\bar{\psi}\psi)^n/(1+\bar{\psi}\psi)^n$
inflation ends once $\bar{\psi}\psi$ reaches $3n/2-1$.  If the initial
value of the scalar bilinear is $(\bar{\psi}{\psi})_i$, then inflation
lasts a number of $e$-foldings $N$ given by
\begin{equation}
  N=\frac{1}{3}\log \frac{(\bar{\psi}{\psi})_i}{(\bar{\psi}{\psi})_e}.
\end{equation}
It follows  that during 60 $e$-foldings  $\bar{\psi}{\psi}$ changes by
eighty  orders  of   magnitude!   This  is  to  be   compared  with  a
conventional chaotic model, where the  scalar field changes by just an
order of magnitude.  This fact  is particularly important in nearly de
Sitter     inflation,     since     the     ``flatness''     condition
(\ref{eq:slow-roll})  has to  be satisfied  for a  range of  values of
$\bar{\psi}{\psi}$ that encompasses eighty orders of magnitude.

An important difference  between inflation driven by a  spinor and the
conventional  scenarios is the  reheating mechanism  after the  end of
inflation.  In  the conventional  scenarios, the universe  is reheated
when  the scalar  field starts  oscillating around  the bottom  of its
potential and decays into particles \cite{Linde, KoLiSt,GreeneKofman}.
If   inflation   is  driven   by   a   spinor   field,  the   quantity
$\bar{\psi}\psi$  evolves  according  to Eq.  (\ref{eq:dilution})  and
hence does not oscillate.   Nevertheless, there are several mechanisms
to  reheat  the  universe.   One  of them  is  gravitational  particle
production at  the end of inflation  \cite{Ford, DamourVilenkin}; more
efficient ways have been suggested in \cite{FeKoLi}.

\subsection{Perturbations}
One  of  the most  appealing  features  of  many of  the  conventional
inflationary models is their  prediction of an adiabatic, nearly scale
invariant spectrum  of primordial density  perturbations, in agreement
with current observations.  Our goal in this section is to compute the
power spectrum  of density perturbations  generated during a  stage of
nearly de  Sitter inflation driven by  the spinor field  $\psi$.  In a
proper  treatment of  the problem,  we would  perturb both  metric and
spinor and solve the linearized Einstein equations.  The nature of the
spinor makes  this path cumbersome, so  we shall rely  on a simplified
analysis,  where  we  only  perturb  the spinor  in  a  given,  fixed,
spacetime background (de Sitter spacetime).

We  shall  characterize  density  perturbations  $\delta\rho$  by  the
variable
\begin{equation}\label{eq:zeta}
  \zeta\equiv \frac{\delta \rho}{\rho+p}.
\end{equation}
This quantity is somewhat analogous  to the Bardeen variable, which is
conserved on large scales in the absence of entropy perturbations, and
which  can be  directly  related to  the  cosmic microwave  background
temperature  fluctuations.  The  source of  the  density perturbations
$\delta\rho$  are the  fluctuations $\delta\psi$  of the  spinor field
around   its  homogeneous  background  value $\psi_0$. 
We treat $\psi_0$ as a classical field, and the fluctuations
$\delta\psi$ as a quantum field in an expanding universe \cite{Parker},
\begin{eqnarray}\label{eq:expansion}
\delta\psi=\frac{1}{(2\pi)^{3/2}}\int d^3k \sum_\sigma
\Big[u(t,\vec{k},\sigma) a(k,\sigma) e^{i \vec{k}\vec{x}}+ \nonumber \\ 
{}+v(t,\vec{k},\sigma) b^\dag(k,\sigma) e^{-i \vec{k}\vec{x}}\Big].
\end{eqnarray}
The index  $\sigma$ runs over the  two spin states of  the spinor, and
the operators  $a$ and $b$ are particle  and antiparticle annihilation
operators,              $\{a(\vec{k},\sigma),a^\dag(\vec{k'},\sigma')\}
=\{b(\vec{k},\sigma),b^\dag(\vec{k'},\sigma')\}
=\delta^{(3)}(\vec{k}-\vec{k'})\delta_{\sigma \sigma'}$.

The power  spectrum $\mathcal{P}(k)$ is a measure  of the fluctuations
of the  variable $\zeta$ on comoving  scales of size $1/k$,  and it is
implicitly defined by the relation \cite{MuFeBr}
\begin{equation}\label{eq:power}
\langle\zeta(t,\vec{x})\zeta(t,\vec{x}+\vec{r})\rangle=
    \int \frac{dk}{k} \frac{\sin kr}{kr} \mathcal{P}(k).
\end{equation}
Here,  $\langle  \,  \rangle$  denotes  the expectation  value  in  an
appropriately chosen vacuum state, $a |0\rangle=b |0\rangle=0$.  Using
expressions (\ref{eq:rho}) and (\ref{eq:p}) for the energy density and
pressure of the spinor field respectively, we find that 
\begin{equation}\label{eq:zeta2}
  \zeta=\frac{\delta\bar{\psi}\, \psi+\bar{\psi}\,\delta\psi}
  {\bar{\psi}\psi},
\end{equation}
where  we  have  dropped  the  subindex $0$  that  denotes  background
quantities.  Substituting Eq. (\ref{eq:zeta2}) into the left hand side
of Eq. (\ref{eq:power}) we obtain (for $\vec{r}=0$)
\begin{equation}
  \langle\zeta(t,\vec{x})\zeta(t,\vec{x})\rangle
  =\frac{2\langle\delta\bar{\psi}(t,\vec{x})\psi(t)\cdot
    \bar{\psi}(t)\delta\psi(t,\vec{x})\rangle}
{(\bar{\psi}\psi)^2},
\end{equation}
where  we have  used the  fact than  only terms  with equal  number of
creation/annihilation  operators   have  a  non-vanishing  expectation
value.

Using  the Pauli-Fierz  identities \cite{Peshkin}  we can  express the
previous  four  spinor  expectation  value in  terms  of  perturbation
bilinears,
\begin{equation}\label{eq:PF}
\langle\zeta\,\zeta\rangle
  =\frac{\langle\delta\bar{\psi}\,\delta\psi\rangle}
{2\bar{\psi}\psi}+
\frac{(\bar{\psi}\,\gamma_a\psi)
  \langle\delta\bar{\psi}\,\gamma^a\delta\psi\rangle}
{2(\bar{\psi}\psi)^2}+\cdots .
\end{equation}
Note that the second term in the right hand side introduces violations
of  isotropy  in the  power  spectrum unless  $\bar{\psi}\gamma_a\psi$
vanishes\footnote{The  power  spectrum  is  isotropic if  the  Fourier
  transform  of   the  correlation  function   on  the  lhs  of   Eq.  
  (\ref{eq:power})  only depends  on $k\equiv|\vec{k}|$  , and  not on
  $\vec{k}$  itself.   For  simplicity,  we  have  implicitly  assumed
  isotropy in  our definition of the power  spectrum $\mathcal{P}$, in
  the rhs of Eq.   (\ref{eq:power}).}.  Because we are only interested
in  an  estimate  of the  amplitude  and  the  $k$ dependence  of  the
spectrum, we can concentrate on the first term on the right hand side.
Substituting  the  expansion (\ref{eq:expansion})  into  that term  we
finally obtain that the power spectrum is of the order
\begin{equation}
  \mathcal{P}(k)\sim
   \frac{k^3}{4 \pi^2}\sum_\sigma
  \frac{\bar{v}(t,\vec{k},\sigma)\,v(t,\vec{k},\sigma)}
    {(\bar{\psi}\psi)}.
\end{equation}

The time  evolution of $v$  is dictated by  the equation of  motion of
$\delta\psi$.  The field  $\delta\psi$ itself satisfies the linearized
Dirac        equation        $i\Gamma^0        D_0\delta\psi+i\Gamma^i
D_i\delta_\psi-m\delta\psi=0$,  where we assume  that $m\equiv  V'$ is
small but non-zero and $V''$  is negligible.  It is convenient to work
with   the   rescaled  field,   $\tilde{\delta\psi}=a^{3/2}\delta\psi$
instead of $\delta\psi$.  The rescaled  field behaves as a spinor with
a time-dependent  mass in flat  space, and in  particular, $\tilde{v}$
satisfies
\begin{equation}\label{eq:linearized}
  i\gamma^0 \frac{d\tilde{v}}{d\eta}
  +\gamma^i k_i \tilde{v}-a m \tilde{v}=0,
\end{equation}
where  $\eta$  denotes conformal  time,  $d\eta=dt/a$.   In de  Sitter
space,   $\eta=-e^{-Ht}/H$   runs  from   ${}-\infty$   to  $0$,   and
$a=-1/(H\eta)$.  Solutions of the Dirac equation (\ref{eq:linearized})
in  a de  Sitter background  were  studied by  Taub \cite{Taub}.   The
ansatz  $\tilde{v}=(v_+V_+,v_-V_-)$,  where $V_+$  and  $V_-$ are  two
time-independent  two-component  spinors,   yields  the  second  order
differential equation
\begin{equation}\label{eq:KG}
  v_\pm''+[k^2+a^2 m^2 \pm i(a m)']v_\pm=0.
\end{equation}
Different linear  combinations of the solutions  to Eq.  (\ref{eq:KG})
correspond  to  different choices  of  vacuum  state.   We choose  the
standard   prescription   where   $v_{\pm}\propto  e^{i   k\eta}$   as
$\eta\to-\infty$  \cite{BirrellDavies}.   The  corresponding  properly
normalized spinor solutions are then
\begin{eqnarray}
  v(\eta,\vec{k},\uparrow)&=&\sqrt\frac{- \pi k \eta}{a^3}\,
  \frac{e^{-\pi m/2H}}{2}\left(
  \begin{smallmatrix}
    H_\nu^{(2)}(-k \eta) k_3/k \\
    H_\nu^{(2)}(-k \eta) (k_1+ik_2)/k \\
    e^{\pi m/H}H_{\bar{\nu}}^{(2)}(-k \eta) \\
    0
  \end{smallmatrix}\right), \nonumber\\
v(\eta,\vec{k},\downarrow)&=&\sqrt\frac{- \pi k \eta}{a^3}\,
\frac{e^{-\pi m/2H}}{2}\left(
  \begin{smallmatrix}
    H_\nu^{(2)}(-k \eta) (k_1-ik_2)/k \\
    -H_\nu^{(2)}(-k \eta) k_3/k \\
    0 \\
    e^{\pi m/H}H_{\bar{\nu}}^{(2)}(-k \eta) \nonumber\\
  \end{smallmatrix}\right).\\ 
\end{eqnarray}
The functions $H_\nu^{(1)}$  and $H_\nu^{(2)}$ are
the   Hankel   functions  of   the   first   kind   and  second   kind
\cite{AbramowitzStegun}, and $\nu=\frac{1}{2}-i m/H$. Up to the factor
$a^{-3/2}$, the previous spinors  oscillate as $e^{i k\eta}$ for modes
inside the horizon, $-k\eta\gg 1$.  Using the asymptotic expansion for
the  Hankel function  in the  limit $-k\eta\ll  1$ (modes  outside the
horizon)
\begin{equation}
  H^{(2)}_\nu (-k\eta)\approx 
  \frac{i}{\pi}\Gamma(\nu)\left(\frac{-k\eta}{2}\right)^{-\nu},
\end{equation}
it is straightforward to verify that the power spectrum ``freezes'' on
large scales and becomes equal to
\begin{equation}\label{eq:powerspectrum}
  \mathcal{P}(k)\sim-\frac{k^3}{2\pi^3 A}\,\left|\Gamma(\nu)\right|^2
  \sinh \frac{\pi m}{H}
\quad\quad (\text{for}\, {}-k \eta\ll 1).
\end{equation} 
Such a power spectrum has spectral index $n=4$, in strong disagreement
with experimental  results consistent with a  scale invariant spectrum
with $n\approx 1$  \cite{flat}. The constant $A$ is  the quantity that
appears in  Eq. (\ref{eq:dilution}). Equation (\ref{eq:powerspectrum})
can  also  be used  to  estimate the  power  spectrum  of the  density
contrast,
\begin{equation}
  \mathcal{P}_{\delta \rho/\rho}\sim 
  \left(\frac{m A}{V a^3}\right)^2\mathcal{P}.
\end{equation}
Because  during  inflation $V$  is  nearly  constant  while $a$  grows
exponentially, spinor fluctuations  are highly suppressed with respect
to, say, scalar field density fluctuations.

The  $k^3$ dependence of  the power  spectrum (\ref{eq:powerspectrum})
and the  $a^3$ decay  of the  density contrast are  to some  extent an
expression  of the  conformal  triviality of  the  system.  Indeed,  a
massless spinor is conformally invariant,  and the power spectrum of a
massless spinor in flat spacetime  displays the same $k^3$ dependence. 
Our  calculation  shows  that   even  the  inclusion  of  a  conformal
symmetry-violating  mass does  not  significantly alter  this result.  
Note  that although  expression (\ref{eq:powerspectrum})  vanishes for
$m=0$, this merely reflects the  chiral asymmetry of the first term in
the expansion (\ref{eq:PF}).  In the limit of zero mass, the discarded
terms give the dominant  contributions, which also are proportional to
$k^3$.

In conclusion, at the level of our simplified preliminary analysis, it
seems that a  stage of (quasi) de Sitter inflation  driven by a spinor
cannot  seed   a  scale  invariant  spectrum   of  primordial  density
perturbations  by itself.   Eventually, a  light scalar  field present
during  inflation (as  in curvaton  models \cite{curvaton})  may solve
this problem.

\section{Non-singular Cyclic Cosmologies}\label{sec:cyclic}
One of the most intriguing issues in cosmology is the ultimate origin
of the universe and the character of its initial state.  
One     of      the     attractions      of     cyclic
cosmologies~\cite{Tolman,BarrowDabrowski,SteinhardtTurok} is that---to
the extent  that they  are truly cyclic,  returning to the  same state
after each cycle---they dispense  altogether with that problem.  Since
they  are  past  eternal,  there  is  no  need  to  formulate  initial
conditions  from  which the  universe  is  evolved  into the  future.  
Furthermore, the universe  has always existed for the  same reason, so
that there is no need to  ask where it originated from.  However, many
of the  cyclic universe models that  have been proposed  so far suffer
from  singularities  that  prevent  a  continuous  account  of  cosmic
history.  At a  certain time,  the universe  evolves into  a singular
state where  the conventional low-energy  effective theory description
of the universe breaks down.  Furthermore, even if the singularity is
regulated in some way, these models can still lead to
inconsistencies~\cite{lev}.

In  this section  we  describe  a scenario  which  avoids this  latter
breakdown (see \cite{branes} for alternatives).  For simple choices of
the self-interaction term $V(\bar{\psi}\psi)$, cyclic cosmologies free
of singularities exist.  Here,  we present a simple model illustrating
this point.  Though  we make no claim that this  simple model leads to
an entirely satisfactory cosmology, we see no obstacle to refining the
basic idea into a more realistic description of our universe.

\begin{figure}
    \includegraphics{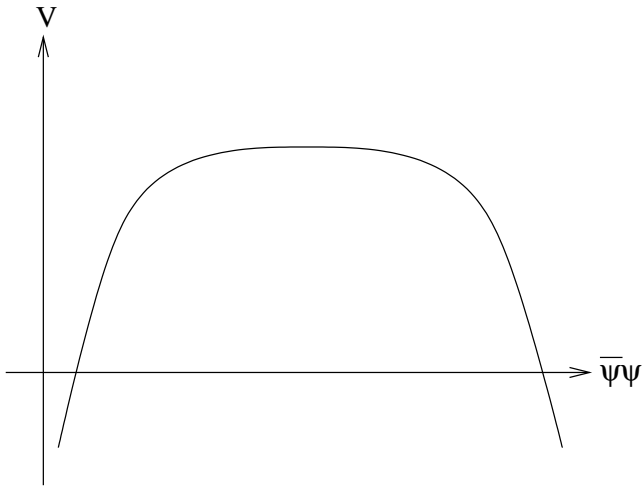}
  \caption
  {Generic form of the potential $V$ in a cyclic non-singular universe.
    \label{fig:V}}
\end{figure}

Consider  a spatially  flat  universe that  contains ``matter''  (dark
energy, dust  and radiation)  and a homogeneous spinor  field $\psi$, 
and suppose
that the  interaction term $V$  in Eq.  (\ref{eq:Lagrangian})  has the
form given  in Figure \ref{fig:V}.  The ``potential''  is negative for
``small''  values of  $\bar{\psi}\psi$,  and it  becomes negative  and
decreases ``fast enough'' for  large values of $\bar{\psi}\psi$.  Such
an interaction might be given for instance by
\begin{equation}\label{eq:model-V}
  V(\bar{\psi}\psi)=\Lambda_\psi+m\bar{\psi}{\psi}-\lambda (\bar{\psi}\psi)^2. 
\end{equation}
Here  $\Lambda_\psi$  is  a   (negative)  contribution  to  the  total
cosmological constant,  $m$ is  a (positive) mass  and $\lambda$  is a
(positive)  coupling  constant.   Hence,  such  a  model  describes  a
conventional,  self-interacting,   massive  spinor  with   a  negative
contribution to the vacuum energy.

\begin{figure*}
    \includegraphics{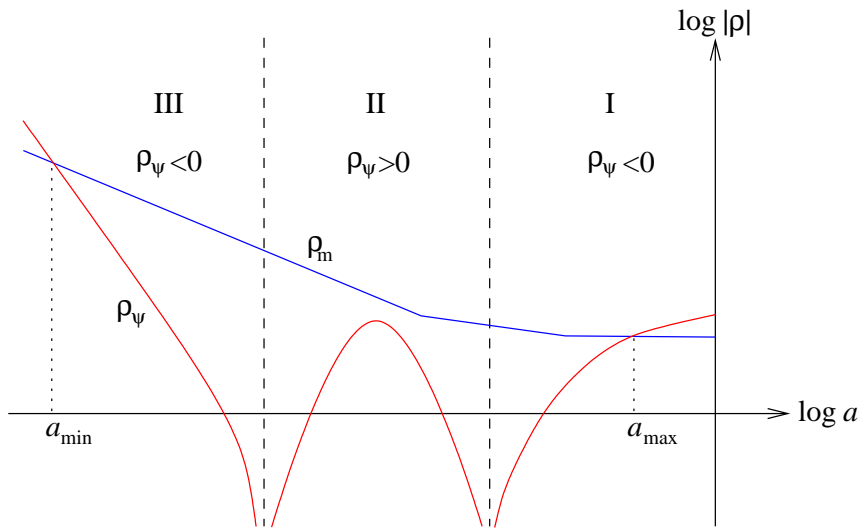}
  \caption
  {Schematic plot of  the energy densities of matter  ($\rho_m$) and 
    spinor ($\rho_\psi$)  as a  function of the  scale factor  $a$. In
    Region I  the energy  density of the  spinor is negative.   If the
    universe expands, it starts  contracting at $a_{max}$, when spinor
    and matter energy densities add  to zero.  In Region II the spinor
    remains  subdominant.  Once  its energy  density  becomes negative
    again  in Region  III,  its importance  starts  growing until  the
    moment  when  the total  energy  density  is  zero again  and  the
    universe bounces back. \label{fig:evolution}}
\end{figure*}

In  order to  describe cosmic  evolution in  such a  universe,  let us
arbitrarily start our description  during expansion.  Suppose that the
matter energy density dominates over  the energy density of the spinor
at  a  time   when  the  latter  is  positive  (region   II  in  Fig.  
\ref{fig:evolution}.)        In       an      expanding       universe
$\bar{\psi}{\psi}\propto a^{-3}$ is driven  to values where the energy
density of the spinor becomes  negative, while the energy densities of
radiation   ($\propto  a^{-4}$)  and   dust  ($\propto   a^{-3}$)  are
``redshifted away'' and  tend to zero. The only  assumption we have to
make at this point is that  the energy density of dark energy does not
increase  at late  times\footnote{By  a suitable  modification of  the
  interaction  in  Fig.   \ref{fig:V},  the spinor  field  could  also
  account  for dark  energy.  In  that case  one can  drop  the latter
  assumption and, up  to the constraints on the  parameters of our toy
  model,  the rest of  our discussion  remains unaltered.}.   Then, if
$\Lambda_\psi$  in  Eq.   (\ref{eq:model-V})  is large  enough,  there
necessarily exists  a value  of the scale  factor $a_{max}$  where the
total energy density $\rho_{tot}$ becomes zero, $\rho_{tot}=0$ (region
I   in   Fig.    \ref{fig:evolution}.)     It   follows   from   Eq.   
(\ref{eq:friedmann}) that at  $a_{max}$, $\dot{a}=0$.  In addition, at
$a_{max}$ the right hand side of of Eq.  (\ref{eq:dotdota}),
\begin{equation}
  \ddot{a}=-\frac{3 a_{max}}{2}\rho_m(w_m-w_\psi),
\end{equation}
is negative, since if $\rho_{tot}$ reaches zero from a positive value,
the combination $w_m-w_\psi$ has  to be positive.  Thus, at $a_{max}$,
$\dot{a}=0$  and $\ddot{a}<0$,  so the  universe  automatically starts
contracting.

After   the  universe  starts   contracting,  $\bar{\psi}{\psi}\propto
a^{-3}$ reverses its motion and  starts growing. At the same time, the
energy  densities in matter  increase as  the universe  contracts.  At
sufficiently small $a$ we can  assume that the matter component of the
universe mainly  consists of radiation.  Assume that  for large values
of $\bar{\psi}\psi$, $V$  decreases faster than the rate  at which the
energy density of radiation  increases, $w_\psi> 1/3$ (recall that the
universe contracts).  If $V$ is  well approximated by a power at large
values    of    $\bar{\psi}\psi$,    this    implies,   from    Eq.    
(\ref{eq:power-potential}),    that    $|V|$    grows   faster    than
$(\bar{\psi}\psi)^{4/3}$,  which  is   satisfied  by  the  interaction
(\ref{eq:model-V}).   Then, the  ratio of  spinor to  radiation energy
densities   steadily   approaches   $-1$   (region   III   in   Fig.   
\ref{fig:evolution}.)   Again,  there   exists  then  a  scale  factor
$a_{min}$  where $\rho_{tot}=0$.  At  $a_{min}$, $\dot{a}=0$  and from
Eq.   (\ref{eq:dotdota}) $\ddot{a}>0$  since, when  $\rho_{tot}=0$ and
$w>1/3$ are satisfied, Eq.  (\ref{eq:dotdota}) reads
\begin{equation}
  \ddot{a}\approx a_{min}\frac{\rho_m}{2}(3 w_\psi-1)>0.
\end{equation}
Hence, at $a_{min}$ the universe bounces and starts expanding until it
again  reaches  region II.   From  that  point  on cosmic  history  as
described  above  repeats  itself.   Note  that  cosmic  evolution  is
singularity-free  throughout.  After  the bounce  from  contraction to
expansion, the  total equation of  state of the universe  evolves from
$w_{tot}=-\infty$ towards  $w_{tot}=1/3$ during radiation  domination. 
Hence, there is a stage  of inflation between the bounce and radiation
domination,  and  eventually  a  spectrum  of  nearly-scale  invariant
perturbations in the matter fields could be seeded.

It   is  possible   to   estimate  the   values   of  the   parameters
$\Lambda_\psi$,  $m$ and $\lambda$  by imposing  certain observational
constraints  on our  model.  Let  us set  today's value  of  the scale
factor to $1$  and denote the present value of  the spinor bilinear by
$(\bar{\psi}\psi)_0$.  Requiring  the energy density of  the spinor to
be subdominant today we find
\begin{equation} \label{eq:c1}
|\Lambda_\psi+m(\bar{\psi}{\psi})_0|\ll \rho_{crit}\approx 10^{-121},
\end{equation}
where  we have  assumed that  the  term proportional  to $\lambda$  is
negligible today.  The previous relation shows that our model requires
a certain degree of fine tuning, but let us point us that the required
fine tuning  is of the  same order as  the one needed to  explain late
time  cosmic acceleration.  If  condition (\ref{eq:c1})  is satisfied,
the spinor field remains subdominant  all the way into the past, until
the moment when its  (negative) energy density exactly compensates the
(positive)  matter  energy density  and  the  universe  bounces (Fig.  
\ref{fig:evolution}).   Primordial  nucleosynthesis  is  the  earliest
epoch when conditions in the universe can be probed.  In order for the
bounce to occur  before nucleosynthesis, and in order  not to conflict
with its standard predictions, the energy density of the spinor should
be subdominant at that time as well.  Since primordial nucleosynthesis
occurs  at $a\approx 10^{-10}$,  and today  $(\rho_r)_0\approx 10^{-4}
\rho_{crit}$, it follows that
\begin{equation}\label{eq:c2}
  \lambda (\bar{\psi}\psi)_0^2 \ll 10^{-145}.
\end{equation}
Finally, in order for our field-theoretic classical description to
remain valid throughout cosmic history, we impose that the
energy density of radiation at the bounce be significantly
below the Planckian energy density, 
\begin{equation}\label{eq:c3}
  \lambda (\bar{\psi}\psi)_0^2\gg 10^{-188}.
\end{equation}
Due  to the  freedom in  the dynamical  variable $(\bar{\psi}\psi)_0$,
there  is a  large  set  of parameters  that  satisfy the  constraints
(\ref{eq:c1}), (\ref{eq:c2})  and (\ref{eq:c3}). It is  easy to verify
that a set of parameters that satisfies all the constraints is
\begin{equation}\label{eq:parameters}
  \Lambda_\psi\approx -10^{-2}\cdot (10^{-3} \text{eV})^4,\, 
  m\approx 10^{-3} \text{eV},\,
  \lambda\approx 10^{-5} \text{GeV}^{-2},
\end{equation}
where  we have  assumed  $(\bar{\psi}\psi)_0=10^{-95}$, which  implies
that at  the bounce $(\bar{\psi}\psi)_{min}\approx  10^{-43}$.  Hence,
throughout cosmic  history $\bar{\psi}\psi$ remains  much smaller than
$1$.  Notice that the cosmological term in (\ref{eq:parameters}) is of
the  same order  as the  component  that is  presently driving  cosmic
acceleration, the  mass $m$ agrees  with common neutrino  mass models,
and  the  coupling constant  $\lambda$  is  of  the order  of  Fermi's
constant.    From  this   point   of  view,   the  parameter   choices
(\ref{eq:parameters}) do not appear to be unphysical.

Certainly,  some aspects  of our  cyclic scenario  still remain  to be
discussed.  We have  not addressed the issue of  how to obliterate the
inhomogeneous debris  of previous cycles,  and we have not  taken into
account particle  production at  the bounce.  Eventually,  both issues
might be related.  Finally, let us  point out that even in the absence
of matter, the  spinor can support a cyclic  universe, where $a_{min}$
and $a_{max}$ are simply  determined by the values of $\bar{\psi}\psi$
where $V$ becomes zero.

\section{Summary and Conclusions}\label{sec:conclusions}
In a  flat FRW universe  there are consistent solutions  of Einstein's
equations coupled to a  homogeneous spinor field.  For such solutions,
the scalar  bilinear $\bar{\psi}\psi$ is proportional  to $a^{-3}$, as
for the  number density of  a gas of non-relativistic  particles.  The
energy density of the spinor is given by an ({\it a priori}) arbitrary
self-interaction  term $V$.   For a  given form  of the  spinor energy
density  $\rho_\psi(a)$,  one   can  always  find  a  self-interaction
$V(\bar{\psi}\psi)$  that  has $\rho_\psi(a)$  as  a  solution of  the
equations of motion.   Thus, canonical, classical, homogeneous spinors
can  violate  any desired  energy  condition,  and  their behavior  in
general  cannot  be reproduced  by  a  minimally coupled,  homogeneous
scalar field.

A  spinor  field  can  also  support  a  sufficiently  long  stage  of
inflation, provided  the self-interaction term $V$  satisfies a single
condition  on   its  slope  for   an  exponentially  large   range  of
$\bar{\psi}\psi$.  This  condition is satisfied, for  instance, if $V$
asymptotes to  a constant value  at large values of  $\bar{\psi}\psi$. 
The spectrum of primordial  spinor density perturbations seeded during
such a stage  has a spectral index $n\approx4$,  and is hence strongly
scale dependent.  In addition,  the power spectrum can be anisotropic,
even though it  is seeded within an FRW-universe.   The existence of a
``curvaton''  field  \cite{curvaton}  during  spinor-driven  inflation
might resolve  these problems, ultimately resulting  in the generation
of   an  adiabatic,   nearly  scale-invariant   spectrum   of  density
perturbations.

Finally, for  simple choices of the self-interaction  $V$, there exist
smooth cyclic  cosmologies where the spinor  energy density oscillates
back  and forth.   The parameters  needed to  accommodate  a realistic
cosmology do not  appear to be unphysical. Although  the simplicity of
the models gets somewhat  distorted, by a straightforward modification
of  the self-interaction  $V$, the  spinor can  also account  for dark
energy and  still allow for  realistic cyclic non-singular  solutions. 

Our approach has  been to treat the spinor field  as a complex valued,
classical object  obeying a simple relativistic equation  of motion, a
non-linear  generalization  of  the  Dirac equation  in  an  expanding
universe.   In  the  Appendix  we  have  addressed  the  validity  and
relevance  of  this assumption.   Additionally,  the  validity of  our
results  certainly  depends  on   the  stability  of  our  homogeneous
solutions against  the growth of inhomogeneous  fluctuations.  We have
left this question for future work.

To conclude, we have shown that spinors can accommodate a large set of
interesting  cosmological solutions.   Although  bilinears generically
decay  during expansion, they  could still  be presently  important if
initially they were sufficiently  displaced out of equilibrium (as for
any  interesting  cosmological solution).   Due  to  its nature,  when
dealing with a spinor field the question to ask is not whether certain
behavior   is  possible,   but  rather,   whether   the  corresponding
self-interaction is natural.

\acknowledgments  It is a  pleasure to  thank Sean  Carroll, Hsin-Chia
Cheng,  Paolo Gondonlo,  Chris Hill,  Slava Mukhanov,  Kazumi Okuyama,
Leonard Parker, and Glenn  Starkman for useful and stimulating remarks
and  discussions.   We especially thank Lev Kofman for useful comments.
CAP  was  supported   by  the  U.S.    DoE  grant
DE-FG02-90ER40560. PBG  was supported  by the DOE  and NASA  grant NAG
5-10842 at Fermilab.

\appendix

\section{Classical spinors}
A  Dirac spinor  is  a four-component  object  $\psi$ that  transforms
according to (\ref{eq:lltspinor}) and  obeys Dirac's equation.  As far
as one  is only concerned with  solutions of equations  of motion, the
components of a  spinor can be consistently regarded  as being complex
numbers,  as  we  have done  in  this  paper.  However, our  world  is
ultimately described by quantum-mechanical  laws, and the question is:
To what  extent is a classical  treatment a good  approximation to the
quantum-mechanical problem?

In  the canonical  approach to  quantum  field theory~\cite{Hatfield},
spinors  are  operator-valued  fields  that act  on  an  appropriately
defined  Hilbert   space.   The  spinor   operator  $\hat{\psi}$  also
satisfies the Dirac equation,
\begin{equation}\label{eq:massiveDirac} 
  i\gamma^\mu \partial_\mu \hat{\psi} -m {\hat{\psi}}=0,
\end{equation}
where for  the purposes of  illustration and simplicity we  consider a
massive  fermion  in  flat  spacetime.   We  work  in  the  Heisenberg
representation,  where  operators are  time-dependent  and states  are
time-independent.  We  would like to  interpret a classical  spinor as
the  expectation   value  of  the  spinor  in   an  appropriate  state
$|s\rangle$,
\begin{equation}
  \psi_{cl}\equiv \langle s|\hat{\psi}|s \rangle
  \equiv\langle\hat{\psi}\rangle.
\end{equation}
Taking the expectation value of equation (\ref{eq:massiveDirac}),
we find that $\psi_{cl}$ satisfies the equation
\begin{equation}
  i\gamma^\mu \partial_\mu\psi_{cl} -m \psi_{cl}=0,
\end{equation}
which simply  states that the  classical spinor $\psi_{cl}$  obeys the
conventional Dirac equation. Therefore,  we already recover one of the
main  ingredients  we  have  used   in  this  paper. 

Note that the expectation value of  a spinor {\em in a physical state}
is  a complex number,  not a  Grassmann number.   There exist  in fact
states  $|c\rangle$  such  that  \mbox{$\langle  c|\hat  \psi_a(x)  |c
  \rangle = \psi_a(x)$}, with  the $\psi_a(x)$'s four Grassmann valued
fields ($a=1,..,4$ is  the spin index). However, that  such states are
not part of the physical Fock  space is easily seen by considering the
energy density in such a state:
\begin{equation}
\rho_c=m\langle c|\bar{\hat\psi}{\hat\psi}|c \rangle= m \bar\psi \psi \, .
\end{equation}
The energy density in a physical state must be a real number.
However, because $\psi_a \psi_b = - \psi_b \psi_a$, $\rho_c^n = 0$ for
any $n > 4$.   This is impossible for a non-zero real number.

Although the expectation value of the spinor obeys the classical Dirac
equation (\ref{eq:massiveDirac}), large quantum fluctuations of $\psi$
around   its  expectation   value  might   invalidate   the  classical
approximation.   In  our particular  case,  the  only observable  that
enters the  classical Einstein equations is the  energy density, which
in our classical treatment is
\begin{equation}\label{eq:classicalrho}
  \rho_{cl}=m\,\bar{\psi}_{cl}\psi_{cl}
\end{equation}
We  want to  find  out whether  the  expectation value  of the  energy
density  $\langle\rho\rangle=m\langle  \bar{\psi}\psi\rangle$ is  well
approximated  by (\ref{eq:classicalrho}).  At this  stage, one  has to
face   a  well-known   problem.  The   vacuum  expectation   value  of
$\bar{\psi}\psi$  is $\sum_k (-1)=-\infty$.   The conventional  way of
dealing with this divergence is to replace expectation values by their
``renormalized'' counterparts,
\begin{equation}
\langle \ldots\rangle_{ren}\equiv \langle s|\ldots|s\rangle- 
\langle 0| \ldots|0\rangle.
\end{equation}
With this  prescription, the expectation value  of $\bar{\psi}\psi$ is
zero for the vacuum, and $n$ for a state containing $n$ particles plus
antiparticles per unit volume.

Then, in order for  our classical approximation to be valid,
the following relation should hold,
\begin{equation}\label{eq:fluctuations}
  \left|\frac{\langle\bar{\psi}\psi\rangle_{ren}
    -\langle\bar{\psi}\rangle_{ren} \langle\psi\rangle_{ren}}
  {\langle\bar{\psi}\rangle_{ren} \langle\psi\rangle_{ren}}\right|
\ll 1.
\end{equation}
In the bosonic case, the  states that satisfy inequalities analogous to
(\ref{eq:fluctuations})  have  large   occupation  numbers.   As  the
largest  occupation number  of fermion  modes is  one, it  is commonly
believed that fermionic physical  states cannot satisfy relations such
as (\ref{eq:fluctuations}).  This turns out not to be the case.

Let $A$ and $B$ be two complex numbers and let $|s\rangle$ be the state
\begin{equation}\label{eq:state}
  |s\rangle=A|0\rangle+B |1\rangle.
\end{equation}
Here, $|0\rangle$ is the vacuum and $|1\rangle=a_0^{\dag}|0\rangle$ is
a zero-momentum one-particle state.  The state is normalized if
\begin{equation}
  |A|^2+|B|^2=1.
\end{equation}
The  spinor operator can be expanded  in creation and
annihilation operators,
\begin{equation} 
 \psi=\sum_{k}
  \left( a_k u_k+  b^\dag{}_k v_k\right),
\end{equation}
where   $u_k$  and  $v_k$   are  normalized   complex-valued  spinors,
$\bar{u}_k  u_k=-\bar{v}_k v_k=1$  and  $\bar{v}_ku_k=\bar{u}_kv_k=0$. 
The reader can easily verify that for the state (\ref{eq:state})
\begin{equation}\label{eq:evs}
  \langle\psi\rangle_{ren}=A^*B u_0,\quad
  \langle\bar{\psi}\rangle_{ren}=B^*A \bar{u}_0, \quad
  \langle\bar{\psi}\psi\rangle_{ren}=|B|^2.
\end{equation}
Therefore, condition (\ref{eq:fluctuations}) implies
\begin{equation}
|B|^2=1-|A|^2\ll 1.
\end{equation}
Obviously,  the last  condition can  be easily  met. It  means  that a
spinor can be treated classically if its quantum state is ``close'' to
the  vacuum.  In  fact,  this is  what  one is  doing  by setting  the
fermions to zero in a  classical treatment of any theory that contains
spinors. But  even if  there are departures  from the vacuum,  we have
shown  that treating  a massive  Dirac spinor  classically is  in some
cases a good approximation.

Finally,  let us  comment on  the fermion  condensates that  are often
encountered  in particle  and  condensed matter  physics.  In  quantum
theories  with self-interacting  fermions,  it might  happen that  the
spinor   bilinear   $\bar{\psi}\psi$   develops  a   non-zero   vacuum
expectation value.  This is what occurs for instance in the BCS theory
of  superconductivity  \cite{BCS},  where phonon-induced  interactions
cause  electrons to  form  bound Cooper  pairs.   In the  relativistic
Nambu-Jona-Lasinio   model  \cite{NambuJona}  or   its  renormalizable
counterpart, the Gross-Neveu model \cite{GrossNeveu}, self-interacting
chiral  fermions  form  a  scalar condensate,  spontaneously  breaking
chiral-symmetry and dynamically generating a fermion mass.  Within the
effective  action  formalism,  the   dynamics  of  the  condensate  is
completely determined  by a classical scalar field  theory.  The exact
classical  theory that  reproduces the  full variety  of  phenomena is
extremely  complicated.  However,  for certain  states of  the quantum
system the effective  action is well approximated by  a simple, local,
relativistic  scalar  field  theory\footnote{Note, however,  that  one
  generically expects  fermion condensates to  couple non-minimally to
  gravity \cite{HillSalopek}.}.  Just  as a strongly coupled fermionic
system  can  be effectively  described  by  a  classical scalar  field
theory, it is conceivable  that certain strongly coupled systems might
be described by a simple classical spinor field theory, as we consider
in this paper.

\end{document}